\documentclass[%
 reprint,
nofootinbib,
 amsmath,amssymb,
 aps,
 prl,
]{revtex4-2}

\usepackage{graphicx}
\usepackage{dcolumn}
\usepackage{bm}

\usepackage[normalem]{ulem}
\usepackage[usenames,dvipsnames]{color}

\begin{document}

\preprint{APS/123-QED}

\title{The Direct Wave is Not a Meaningful Test of Horizon Properties}

\author{Anuj Kankani}
\email{anuj.kankani@mail.wvu.edu}
\affiliation{Department of Physics and Astronomy, West Virginia University, Morgantown, WV 26506, USA
}
\affiliation{Center for Gravitational Waves and Cosmology, West Virginia University, Chestnut Ridge Research Building, Morgantown, WV 26505,
USA
}
\author{Sean T. McWilliams}
\affiliation{Department of Physics and Astronomy, West Virginia University, Morgantown, WV 26506, USA
}
\affiliation{Center for Gravitational Waves and Cosmology, West Virginia University, Chestnut Ridge Research Building, Morgantown, WV 26505,
USA
}

\date{\today}

\begin{abstract}
Recently, a distinct non--quasinormal mode component of black hole binary merger radiation, named the direct wave, has been identified. The frequency and damping time of the direct wave have been associated with properties of the remnant horizon. This has led to direct-wave based analysis of GW250114, including a test of Hawking's area law. However, as we demonstrate here using numerical relativity strain data, the direct wave frequency is not correlated with the horizon frequency or surface gravity, other than an incidental crossing around $\chi_f \approx 0.7$ corresponding closely to the remnant spin of GW250114. We show that while the instantaneous frequency of the direct wave is quasi-stable, the damping time shows significant evolution and therefore a single damped sinusoid model, containing a fixed damping time, is not appropriate. We further show that an evolving frequency model based on horizon properties also does not model the direct wave component for systems with large remnant spins. We demonstrate that testing Hawking's area law with a horizon frequency based on the direct wave interpretation will lead to apparent violations of Hawking's area law when no violation actually occurs. Our results therefore indicate that the direct wave is not a reliable probe of the remnant horizon's properties.
\end{abstract}

\maketitle
\paragraph{Introduction.---}
With the ever increasing number of gravitational waves (GW) detected by LIGO \cite{abac2025gwtc,abac2025gwtc_methods}, along with the expected increased sensitivity of future GW detectors such as LISA \cite{LISA}, Cosmic Explorer \cite{CE} and the Einstein Telescope \cite{Einstein_tel}, there has been significant focus on understanding the merger-ringdown gravitational radiation emitted from a binary black hole (BBH) merger. Significant work is focused on using gravitational waves to perform strong field tests of General Relativity \cite{grtest1,grtest2,grtest3,grtest4,grtest5,chung2026measuringblackholesarea, berti2025black}. One common approach is through the detection of multiple quasinormal modes (QNMs), the characteristic frequencies and damping times associated with the perturbed BH emerging from a BBH merger \cite{berti2009quasinormal}. Reliably detecting multiple QNMs requires understanding when the gravitational wave can be described by a sum of linear QNMs \cite{giesler2019black,cheung2024extracting,pacilio2024flexible,baibhav2023agnostic,magana2025high}.

Through the use of rational filters \cite{ma2022quasinormal,ma2023using,lu2025statistical}, frequency domain filters designed to target QNM frequencies, Oshita \cite{oshita2025probing} identified a distinct non--QNM portion of the merger radiation, which they dubbed the direct wave \cite{oshita2025probing,lu2025gw250114,kankani2026direct,chung2026measuringblackholesarea,dyer2026modelingdirectwavesbinary}. The direct wave has been associated with the prompt emission of the plunging perturber \cite{oshita2025probing, lu2025gw250114, chung2026measuringblackholesarea}. In the model developed in \cite{oshita2025probing}, for high remnant spin systems, the direct wave is expected to oscillate around, but not exactly at, the horizon frequency of the remnant, and the damping time is expected to be characterized by the surface gravity of the remnant. The authors of \cite{oshita2025probing} showed this in the comparable mass limit through the application of rational filters to SXS:BBH:0305 \cite{SXS:BBH:0305,SXSCatalogData_3.0.0,SXSPackage_v2026.0.0,sxs_cat3}, a GW150914 \cite{gw150914} like system. If the direct wave were directly linked with the horizon properties of the remnant black hole, it could provide a complementary test to QNM based tests of the remnant black hole. Due to its high SNR, GW250114 \cite{abac2025gw250114} has become the primary target for identifying the direct wave in LIGO data. The authors of \cite{lu2025gw250114} found evidence of the direct wave in GW250114, with a frequency and damping time consistent with the horizon frequency and surface gravity of the remnant black hole. In \cite{chung2026measuringblackholesarea}, GW250114 was used to perform a test of Hawking's area law, providing a complementary test to those done with standard ringdown analysis. 

However, as we originally pointed out for the gravitational wave News \cite{kankani2026direct}, using both NR data and the Backwards One Body model \cite{mcwilliams2019analytical,kankani2025bob,kankani2025modeling,mahesh2025combining}, and more recently suggested by \cite{dyer2026modelingdirectwavesbinary}\footnote{We note that \cite{dyer2026modelingdirectwavesbinary} appeared during the final preparation of this manuscript and did not influence its conclusions, so the results provide independent refutations of \cite{lu2025gw250114,chung2026measuringblackholesarea}.}, the instantaneous direct wave frequency does not oscillate near the horizon frequency across the parameter space. The instantaneous direct wave frequency only incidentally coincides with the horizon frequency at $\chi_f \approx 0.7$. Even for high remnant spin systems, the direct wave frequency is clearly separated from the horizon frequency. SXS:BBH:0305, has a final spin of $\chi_f \approx 0.69$ and GW250114 has a final spin near 0.68 \cite{abac2025gw250114}. Both of these systems therefore correspond to remnant parameters where the direct wave frequency \textit{coincidentally} occurs near the horizon frequency. 

In this work, we expand on this finding using NR data for the gravitational wave strain. We show that direct wave models based on the horizon frequency and surface gravity of the remnant are not physically applicable for the direct wave, even for highly spinning remnants. We further show that incorrectly assuming the direct wave is linked with the horizon frequency and surface gravity can lead to apparent violations of Hawking's area law. We emphasize that because GW250114 has a remnant spin where the direct wave frequency happens to be near the horizon frequency, the extracted frequencies of the direct wave in \cite{lu2025gw250114} and \cite{chung2026measuringblackholesarea} are likely correct. However, it is the claim of an association of these frequencies with the remnant horizon, and any subsequent interpretation, that needs to be considered carefully.

\paragraph{Applying Rational Filters.---}
In order to better understand the evolution of the direct wave, we apply rational filters \cite{ma2022quasinormal,ma2023using,lu2025statistical} to a selection of quasicircular and non-precessing SXS data \cite{sxs_cat1,sxs_cat2,sxs_cat3}. Following \cite{oshita2025probing}, throughout this work our rational filter will be composed of 7 prograde QNMs, 2 retrograde QNMs and the $(l=3, m=2, n=0)$ prograde QNM necessary to remove the spherical-spheroidal mixing present in SXS data \cite{ma2022quasinormal}. Our analysis will focus solely on the $(2,2)$ mode of the gravitational wave strain. Additional information on and analysis of these filters is detailed in \cite{ma2022quasinormal, oshita2025probing}. 

In \cite{oshita2025probing}, SXS:BBH:0305 was used as evidence to show that the instantaneous direct wave frequency oscillates near twice the horizon frequency, $\Omega_H = \chi/(2r_+)$, and the damping is characterized by the surface gravity, $\kappa = \sqrt{1-\chi^2}/(2r_+)$, of the remnant black hole. Here $\chi$ is the dimensionless remnant spin magnitude of the black hole and $r_+ = 1 + \sqrt{1 - \chi^2}$ is the radius of the outer Kerr event horizon. We emphasize that due to the possible screening of the horizon mode, the authors of \cite{oshita2025probing} only state that the direct wave should oscillate near the horizon frequency. Furthermore, they argue that the damping rate is only characterized by the surface gravity, $-\mathrm{Im} (\omega) \sim \mathcal{O}(\kappa) $. However, as we show in this work, the direct wave is not correlated with the horizon frequency and $\mathrm{Im}(\omega)$ significantly varies over a short interval of time.

Since the direct wave-horizon correlation primarily stems from the frame dragging induced by the remnant space time, we focus on an extremely high remnant spin SXS system, SXS:BBH:0178 \cite{sxs_cat3,SXSCatalogData_3.0.0,SXSPackage_v2026.0.0,Scheel_2015,SXS:BBH:0178}, with remnant spin $\chi_f \approx 0.95$. This simulation is available at multiple resolution levels and in the supplemental materials we show that our results are not limited by the numerical resolution nor by insufficient QNMs in our rational filter. Throughout this work we will model the filtered waveform as a damped sinusoid with either a constant or time evolving complex frequency $\omega(t) = \omega_R (t) + i\omega_I (t)$. $\mathrm{Re}(\omega)$ represents the instantaneous frequency and the damping time is given by $\tau (t) \equiv -1/\omega_I (t)$. Because we focus on the (2,2) mode, using $\omega = m\Omega$, we will compare the instantaneous frequency to $2\Omega_H$.

\begin{figure}
    \centering
    \includegraphics[]{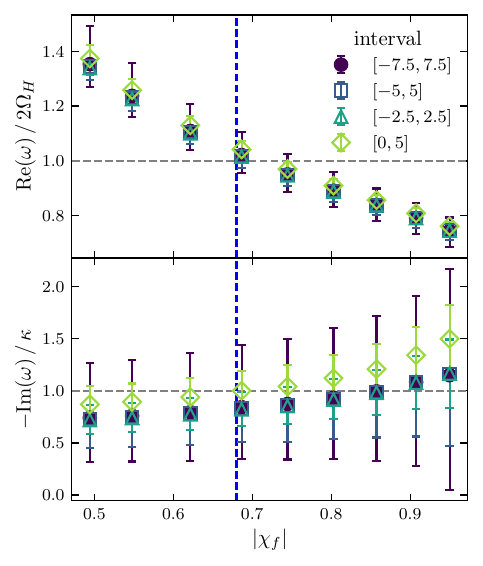}
    \caption{Top: Ratio of the real part of the direct wave frequency to twice the horizon frequency for various time intervals. The different styles of scatter points represent the average value and the error bars indicate the maximum and minimum values over the respective intervals. We use a selection of quasicircular, non-precessing, and equal mass SXS simulations with varying remnant spin $\chi_f$. Bottom: Same as the top panel but for the ratio of the imaginary part of the direct wave frequency to the surface gravity. In both panels, the vertical (blue) dashed lines show $\chi_f$ corresponding to GW250114.}
    \label{fig:limited_sxs_cases}
\end{figure}
To demonstrate that the direct wave frequency is not related to the remnant properties of the horizon, in Fig.~\ref{fig:limited_sxs_cases} we use a selection of equal mass, quasicircular, and non-precessing systems from the SXS catalog. These systems were chosen based on their remnant spin values and available resolution levels. We focus on systems with $\chi_f \gtrsim 0.5$ and compare the extracted direct wave frequency with the horizon frequency and surface gravity. We compute the average value of the direct wave frequency over four different time intervals around the merger time. Our error bars indicate the maximum and minimum values of the direct wave frequency over the respective interval. 

We clearly see that the real part of the direct wave frequency is much larger than $2\Omega_H$ for smaller remnant spins, and becomes much smaller than $2\Omega_H$ as we increase the remnant spin, so clearly does not track the horizon frequency. The average value agrees well over all intervals and the maximum and minimum values do not significantly deviate from the average value for higher remnant spins, indicating the quasi-stable nature of the instantaneous frequency. 

The average value of $-\mathrm{Im}(\omega)$ for all four intervals similarly does not track the surface gravity, and there is a significant difference between the maximum, minimum, and average values, indicating that the damping time is going through significant evolution even over these brief intervals. The range of $-\mathrm{Im}(\omega)$ also includes the values of several prograde QNMs. Therefore, we cannot say that $-\mathrm{Im}(\omega)$ is physically related to the surface gravity $\kappa$ any more than it is related to any other physical damping time associated with the remnant. Furthermore, the significant range of $-\mathrm{Im}(\omega)$ is important to account for when performing any joint QNM-direct wave analysis.
    
\begin{figure}
    \centering
    \includegraphics[]{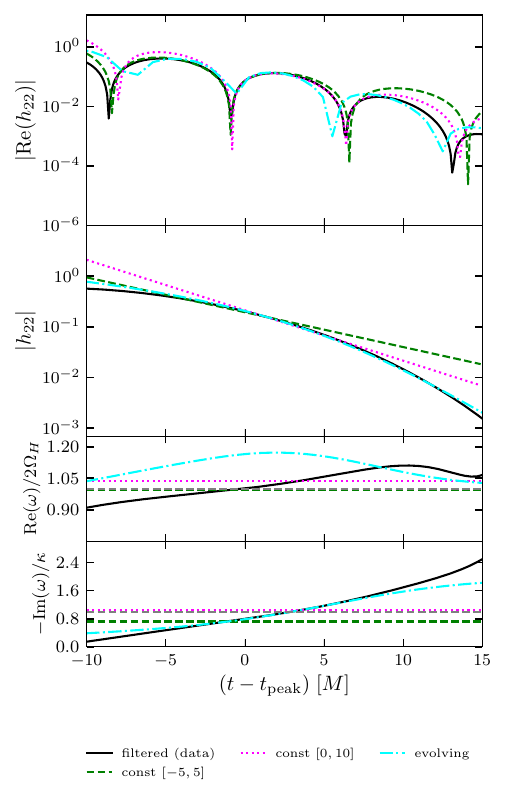}
    \caption{Analysis of SXS:BBH:0305, a GW150914 like system with remnant spin $\chi_f = 0.69$. First panel: Absolute value of the real part of the filtered strain. Second panel: Absolute value of the full filtered strain. Third panel: Ratio of the real part of the filtered frequency to twice the horizon frequency. Lower panel: Ratio of the imaginary part of the filtered frequency to the surface gravity. In solid black we show the NR filtered data. In dashed green and dotted red we show the result from fitting a constant frequency over two different time intervals. In dash-dotted cyan we show the result from the time dependent evolving frequency model described in \cite{oshita2025probing}. $t_{\rm{peak}}$ refers to the time of the peak amplitude of the (2,2) mode of the unfiltered NR strain. All time intervals are given in units of $M$.}
    \label{fig:single_fit_0305}
\end{figure}

\begin{figure}
    \centering
    \includegraphics[]{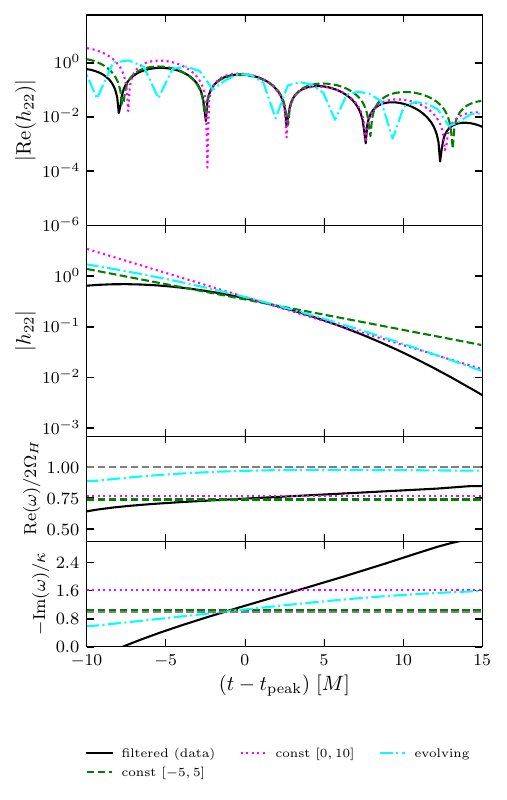}
    \caption{Analysis of SXS:BBH:0178, a system with remnant spin $\chi_f \approx 0.95$. All panels and line styles are the same as in Fig.~\ref{fig:single_fit_0305}.}
    \label{fig:single_fit_0178}
\end{figure}

\paragraph{Modeling the Direct Wave.---}
Two main methods have been used to model the direct wave. The authors of \cite{lu2025gw250114,chung2026measuringblackholesarea,dyer2026modelingdirectwavesbinary} used a damped sinusoid model with a constant complex frequency, while the authors of \cite{oshita2025probing,lu2025gw250114} used linear black hole perturbation theory to develop a model based on the direct wave-horizon interpretation. Critically, this latter model allows for a time-dependent frequency evolution:
\begin{equation}
    h_{DW}(t) \propto \frac{[(\omega(t) - 2\Omega_H) + i\kappa][(\omega(t) - 2\Omega_H) + 2i\kappa]}{(2i\omega(t))(1+\beta)} e^{-i\phi(t)}
\end{equation}
where $\phi(t) = \int\omega(t)(1+\beta)dt$, $\beta \equiv -dx/dt$ is the local velocity, and $\omega(t)$ is determined through the Kerr geodesic equations.
A more detailed description of this model can be found in \cite{oshita2025probing, lu2025gw250114}. In Figs.~\ref{fig:single_fit_0305} and \ref{fig:single_fit_0178} we show the results of both models. For the constant frequency model, we fit a complex amplitude and a complex frequency to the filtered waveform over two different time intervals. We align the evolving frequency model such that $-\mathrm{Im}(\omega)/\kappa = 1$ at the same time as the filtered data and allow for an overall normalization with the NR data. We show that both constant frequency fits recover similar values for $\mathrm{Re}(\omega)$, very close to twice the horizon frequency. This is expected since, as noted in \cite{kankani2026direct}, there is an incidental crossing between the direct wave frequency and the horizon frequency where the remnant spin of SXS:BBH:0305 lands. The evolving model deviates slightly further from, but eventually asymptotes to, the horizon frequency, replicating what is seen in \cite{oshita2025probing}. The two constant frequency fits are unable to track the evolving $-\mathrm{Im}(\omega)$, resulting in them only being valid over very short time intervals where the direct wave signal has not had sufficient time to damp significantly. The evolving model for this case, replicating the result of \cite{oshita2025probing}, does a much better job of tracking the frequency.

However, Fig.~\ref{fig:limited_sxs_cases} along with the results of \cite{kankani2026direct} suggest that if we go to a higher remnant spin, neither the constant nor the evolving frequency models based on the horizon frequency or surface gravity of the remnant should be accurate. In Fig.~\ref{fig:single_fit_0178} we apply the rational filter to SXS:BBH:0178, a $\chi_f \approx 0.95$ system, and use both the constant and evolving frequency models. Immediately we can see that the horizon frequency based evolving frequency model used in \cite{oshita2025probing,lu2025gw250114} does not accurately model the direct wave. Because the evolving frequency model is based on the direct wave-horizon interpretation, and the actual instantaneous direct wave frequency for this high remnant spin system lies significantly lower, it models the incorrect frequency. For the imaginary component, the model does not evolve as rapidly as the NR direct wave imaginary frequency. The net result of this incorrect modeling of the frequency is seen in the top panel of Fig.~\ref{fig:single_fit_0178}, where the evolving frequency model differs significantly from the NR direct wave result. The constant frequency models, which treat the complex frequency as a free parameter, are able to better model the direct wave because their recovered values for $\mathrm{Re}(\omega)$ differ by approximately 25\% from the horizon frequency. For the imaginary component, the two fitting intervals differ significantly in their recovered value because, as seen in the bottom panel of Fig.~\ref{fig:single_fit_0178} the NR direct wave value is evolving significantly over a short period of time. However, similar to the SXS:BBH:0305 case, if the direct wave is being modeled over sufficiently short intervals, where there is not enough time for the direct wave to damp significantly, then a constant frequency model can adequately model the direct wave. 

\begin{figure}[h]
    \centering
    \includegraphics[]{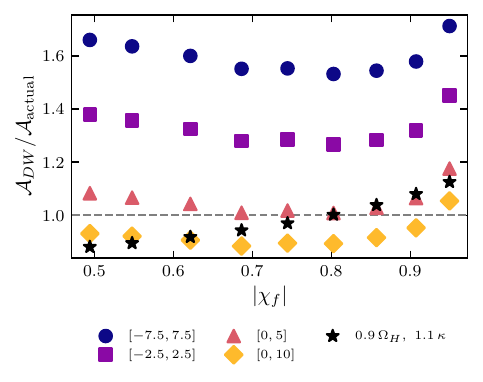}
    \caption{Ratio of the direct wave calculated horizon area, $\mathcal{A}_{DW}$ to the horizon area calculated using the NR final mass and spin $\mathcal{A}_{\mathrm{actual}}$. We show the results for varying remnant spin and fitting time intervals used to obtain the direct wave frequency. All time intervals are in units of $M$ and $t=0$ refers to the time of the peak amplitude of the (2,2) mode of the unfiltered NR strain. We also show the area obtained by 10\% deviations of the horizon properties of the remnant (black star).}
    \label{fig:hawking}
\end{figure}

\paragraph{Testing Hawking's area law.---}
If the direct wave were a reliable method of obtaining the horizon frequency and surface gravity of the remnant black hole, then it would provide a complementary test of Hawking's area law to those done with standard ringdown analysis. As shown in \cite{chung2026measuringblackholesarea}, the horizon area can be written in terms of $\Omega_H$ and $\kappa$ as
\begin{equation}
    \mathcal{A}_{DW} = \frac{2\pi}{\sqrt{\Omega_H^2 + \kappa^2}\bigg(\sqrt{\Omega_H^2 + \kappa^2} + \kappa\bigg)}.
    \label{eq:area}
\end{equation}
The authors of \cite{chung2026measuringblackholesarea} used a constant frequency model to test Hawking's area law by identifying the direct wave in GW250114. However, as we have noted several times, GW250114 has a remnant spin of $\chi_f \approx 0.68$, exactly where the instantaneous direct wave frequency \textit{coincidentally} matches the horizon frequency. As seen in Fig.~\ref{fig:limited_sxs_cases}, this relationship does not hold as we go to higher or lower remnant spins. As seen in Fig.~\ref{fig:single_fit_0178}, a constant frequency model will recover values significantly different from the actual horizon parameters for high remnant spin systems, contrary to the expectation that the model should perform better for higher spins. Therefore, using the direct wave to test Hawking's area law for systems with $\chi_f$ away from 0.7, may lead to an apparent violation of Hawking's area law. But such a violation would not be physical; it would occur due to the incorrect association of the recovered frequency with the horizon parameters of the remnant. In Fig.~\ref{fig:hawking} we show the ratio of the recovered $\mathcal{A}_{DW}$ for various fitting intervals to the actual horizon area, calculated using the NR remnant mass and spin values, for a selection of SXS equal mass, quasicircular and nonprecessing systems with $\chi_f \gtrsim 0.5$. As can be seen, various fitting intervals result in $\mathcal{A}_{DW}$ values that are inconsistent with the true remnant horizon area, as computed from the NR remnant mass and spin. A direct wave based test of Hawking's area law would therefore report apparent violations that are not physical. Only between spins of $0.65 \lesssim |\chi_f| \lesssim 0.85$ and a tight fitting window of $[0M,5M]$ do we obtain a result consistent with the area as determined from the remnant mass and spin of the NR simulation. However, even here we note that care has to be taken because, as seen in the black star data points in Fig.~\ref{fig:hawking}, if $\Omega_H$ and $\kappa$ deviate in opposite directions, it is still possible to recover a consistent $\mathcal{A}_{DW}$. Therefore even a horizon frequency based test that is consistent with Hawking's area law may recover frequencies that deviate from the actual horizon parameters.

\paragraph{Discussion.---}
Using rational filters applied to numerical relativity strain data, we have investigated the direct wave for remnant spins $|\chi_f| \gtrsim 0.5$. We find that the instantaneous direct wave frequency does not oscillate around $2\Omega_H$, and the direct wave damping likewise does not track the surface gravity $\kappa$ and in fact evolves rapidly and spans a range that is far too wide to be meaningfully characterized by $\kappa$.

These findings have important consequences for tests of general relativity. If the direct wave were a faithful probe of the horizon frequency and surface gravity, the horizon-based evolving frequency model used in \cite{oshita2025probing,lu2025gw250114} would recover the horizon properties more reliably as we go to higher remnant spins. However, we find that the deviations increase with remnant spin, suggesting that the direct wave is not a useful probe of the horizon frequency and surface gravity. Therefore, as we show in this work, associating the recovered direct wave frequency with the horizon parameters can lead to apparent violations of Hawking's area law that are due purely to the physically incorrect model being applied. Lastly, we stress that our results only apply to the comparable mass cases we are considering. The authors of \cite{oshita2025probing} studied the direct wave for both extreme mass ratio inspirals (EMRIs) and comparable mass cases. It is possible that the direct wave -- horizon interpretation is still valid in the EMRI limit, but further investigation is warranted. While the direct wave will be a crucial part of understanding the structure of merger-ringdown radiation and improving QNM fitting, it is critical that we identify the correct physics driving the direct wave in order to gain meaningful physical insights.

\paragraph{Acknowledgments.---}
AK and STM were supported in part by NASA grants 22-LPS22-0022 and 24-2024EPSCoR-0010. 
\appendix
\section*{Supplemental Material} 
\begin{figure}
    \centering
    \includegraphics[]{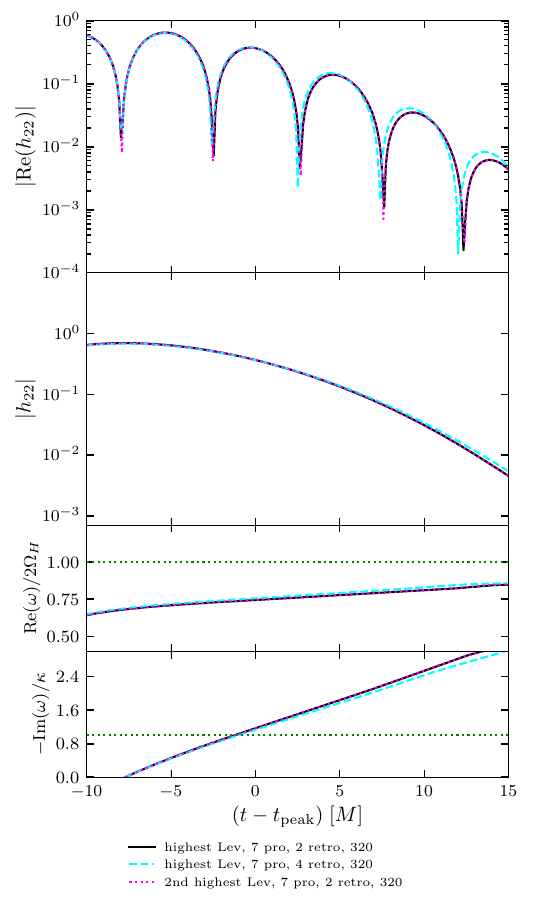}
    \caption{Comparison of the highest (solid black) and second highest (dotted magenta) resolutions for SXS:BBH:0178. We also show the highest level resolution but with two additional retrograde modes removed (dashed cyan). All waveforms are time- and phase-aligned.}
    \label{fig:systematics}
\end{figure}

Since SXS:BBH:0178 is an extremely high spin remnant system, in Fig.~\ref{fig:systematics} we show the two highest resolution levels, as well as a third curve where we have removed two additional retrograde modes on top of our standard 7 prograde, 2 retrograde and $(l=3, m=2, n=0)$ prograde modes. All three curves are time- and phase-aligned and show excellent agreement with each other, indicating that numerical noise is not affecting our results and we are removing a sufficient number of QNMs from our waveform.
In Table~\ref{tab:sxs_cases}, we list the SXS simulations used in this work along with their remnant spins.
\begin{table}
    \centering
    \begin{tabular}{l@{\hspace{1cm}}c}
        \hline
        SXS ID & $\chi_f$ \\
        \hline
        SXS:BBH:2089 & 0.4942 \\
        SXS:BBH:2088 & 0.5478 \\
        SXS:BBH:3919 & 0.6215 \\
        SXS:BBH:4434 & 0.6864 \\
        SXS:BBH:0305 & 0.6921 \\
        SXS:BBH:2095 & 0.7447 \\
        SXS:BBH:3927 & 0.8029 \\
        SXS:BBH:2099 & 0.8574 \\
        SXS:BBH:3897 & 0.9075 \\
        SXS:BBH:0178 & 0.9499 \\
        \hline
    \end{tabular}
    \caption{Remnant dimensionless spin $\chi_f$ for each SXS case considered in this work.}
    \label{tab:sxs_cases}
\end{table}


\bibliography{citations}

\end{document}